\documentclass[conference]{IEEEtran}
\IEEEoverridecommandlockouts
\usepackage{cite}
\usepackage{caption} 
\usepackage{amsmath,amssymb,amsfonts}
\usepackage{algorithmic}
\usepackage{pdfpages}
\usepackage{graphicx}
\usepackage{booktabs}
\usepackage{textcomp}
\usepackage{xcolor}
\usepackage{tablefootnote}
\usepackage[misc]{ifsym}
\usepackage{subfig}
\usepackage{adjustbox}
\usepackage{amsmath,amssymb,amsfonts}
\usepackage{graphicx}
\usepackage{amsthm}
\usepackage{multirow}
\usepackage{bibentry}
\usepackage{listings} % Source code formatting and highlighting
\usepackage{float}
\usepackage{color}
\usepackage{url}
\usepackage[ruled,linesnumbered]{algorithm2e}

\ifCLASSOPTIONcompsoc

  \usepackage[nocompress]{cite}
\else

  \usepackage{cite}
\fi

\ifCLASSINFOpdf

\else

\fi

\newtheoremstyle{custom}
  {\topsep}   % Space above
  {\topsep}   % Space below
  {\itshape}  % Body font
  {}          % Indent amount
  {\bfseries} % Theorem head font
  {:}         % Punctuation after theorem head
  {.5em}      % Space after theorem head
  {}          % Theorem head spec

\theoremstyle{custom}


\renewcommand{\proofname}{Proof}

\begin{document}

\title{SplitLLM: Hierarchical Split Learning for Large Language Model over Wireless Network
%\vspace{-0.6cm}
}

\author{\IEEEauthorblockN{
Songge Zhang$^{\star,*}$, %Conghao Zhou$^\star$, 
     Guoliang Cheng$^{\star,\dagger}$, % Qiang Ye$^\diamond$, 
         Zuguang Li$^{\star,\ddagger}$,
        and
      Wen Wu$^{\star\textrm{,\;\Letter}}$}
    % School of Electronic and Computer Engineering
        \IEEEauthorblockA{
    Frontier Research Center, Peng Cheng Laboratory, Shenzhen, China$^\star$\\
    Peking University Shenzhen Graduate School, Peking University, Shenzhen, China$^*$\\
  The Department of Electronic and Electrical Engineering, Southern University of Science and Technology$^\dagger$\\
 Department of Electronics and Information Engineering, Harbin Institute of Technology (Shenzhen), Shenzhen, China$^\ddagger$\\
     Email: \mbox{\{zhangsg, chenggl, lizg01, wuw02\}@pcl.ac.cn}
    }
    \vspace{-0.8cm}

}

    %\vspace{-0.85cm}

\maketitle
\begingroup\renewcommand\thefootnote{${\textrm{\Letter}}$}
\footnotetext{Wen Wu (wuw02@pcl.ac.cn) is the corresponding author of this paper.
}

\endgroup
% \begingroup\renewcommand\thefootnote{${\textrm{\Letter}}$}
% \footnotetext{Guoliang Cheng (wuw02@pcl.ac.cn) is the corresponding author of this paper.}
% \endgroup
\begin{abstract}

Fine-tuning a large language model (LLM) using the local data of edge users can enable personalized services and applications. For privacy protection, the prevalent solution adopts distributed learning for fine-tuning and integrates low-rank adaptation (LoRA) to reduce users' computational load. However, as the number of users increases, numerous users simultaneously communicate with the server, and multiple server-side models concurrently execute on the server, leading to significant communication congestion and memory pressure.
In this paper, we propose a split learning (SL) scheme for fine-tuning LLM in wireless networks, which involves one cloud server, a small number of edge servers, and multiple users. 
Specifically, the pre-trained model and LoRA adapters are divided into three parts and deployed across the cloud, edge, and user sides. The training process follows the sequence of user, edge, and cloud, with forward and backward propagation achieved by transmitting activation and gradient. In each round, all edge servers and an equivalent number of users train in parallel, and only the LoRA adapters are updated. At the end of each round, all edge-side and user-side LoRA adapters are uploaded to the cloud for aggregation.
Extensive simulation demonstrates that the proposed scheme can reduce peak memory usage up to $74\%$ compared to the state-of-the-art benchmarks.
\end{abstract}

\section{Introduction}

% 1、介绍LLM火热；2、2-3句话，LLM基于公共数据训练，边缘端分布很多用户数据，需要这些数据去个性化fine-tuning LLM；3、LLM需要fine-tuning，fine-tuning带来的好处。 

With great capabilities in tackling complex tasks, large language model (LLM) has enabled a wide range of real-world applications across various domains, such as natural language processing,  computer vision, and artificial
general intelligence~\cite{kuang2023federatedscope, driess2023palm-e}. The effectiveness of LLM indeed hinges significantly on both the size of the models and the breadth of their training datasets~\cite{chen2023federated}. LLM can capture more complex patterns and nuances in language, while extensive datasets provide the diverse examples needed to train these models effectively. However, public domain data can sometimes lack the depth, diversity, and specificity needed for training LLM~\cite{lin2024data}. With the advancements in sensing technology, edge devices are increasingly capable of generating and collecting vast amounts of data, which can be leveraged for customized LLM fine-tuning. The LLM fine-tuning gives its capability of reducing the number of trainable parameters for pre-trained LLM while achieving comparable learning performance with the full-model fine-tuning~\cite{wen2023hard}.

Significant research effort has been devoted to improving fine-tuning schemes for enhancing the performance of LLM. To protect privacy, distributed training paradigms such as federated learning (FL) can be applied to fine-tuning~\cite{babakniya2023slora}. In~\cite{kalapaaking2022blockchain,tran2019federated, han2021fedmes}, the FL-based schemes enable a public server and multiple local users to train a model collaboratively without directly sharing local data. In~\cite{sheng2023sLoRA,hu2021lora}, low-rank adaptation (LoRA) is applied to reduce communication, memory, and computation costs in the training process. LoRA can update the model by decomposing the incremental matrix of attention weight into low-rank matrices, reducing the number of model parameters. In~\cite{lin2024splitlora}, a split learning (SL)-based fine-tuning scheme is proposed, which offloads the primary training workload to a server via model partitioning. It exchanges activation between user and server to reduce communication overhead, as the activation has smaller data sizes compared to the entire pre-trained model, such as GPT-2~\cite{radford2019language} and LLaMA-1~\cite{fang2024automated}.

Despite these works enabling the possibility of fine-tuning at the edge, they don’t consider the scenario with a large number of users and still face memory pressure. First, existing work assumes a cloud or an edge computing platform as the server~\cite{liu2020client, cui2022optimizing}. In cloud-based schemes, the large number of participating users, potentially reaching millions, leads to significant network congestion and inefficiency in the training process~\cite{wu2020fedhome,bonawitz2019towards}. In edge-based schemes, each server can access only a limited number of users, resulting in inevitable training performance loss~\cite{wang2021edge}. Second, while existing works can reduce users' memory usage through LoRA, they cause additional memory pressure on the server. The reason is that as the number of users increases, the server needs to execute multiple server-side model training in parallel~\cite{lin2024splitlora}. For the above reasons, a new scheme is needed to support LLM fine-tuning and accommodate large-scale users and limited server memory capacity.

\begin{figure*}[htbp]
\centering
\subfloat[Proposed SplitLLM]{\includegraphics[width=2.8in]{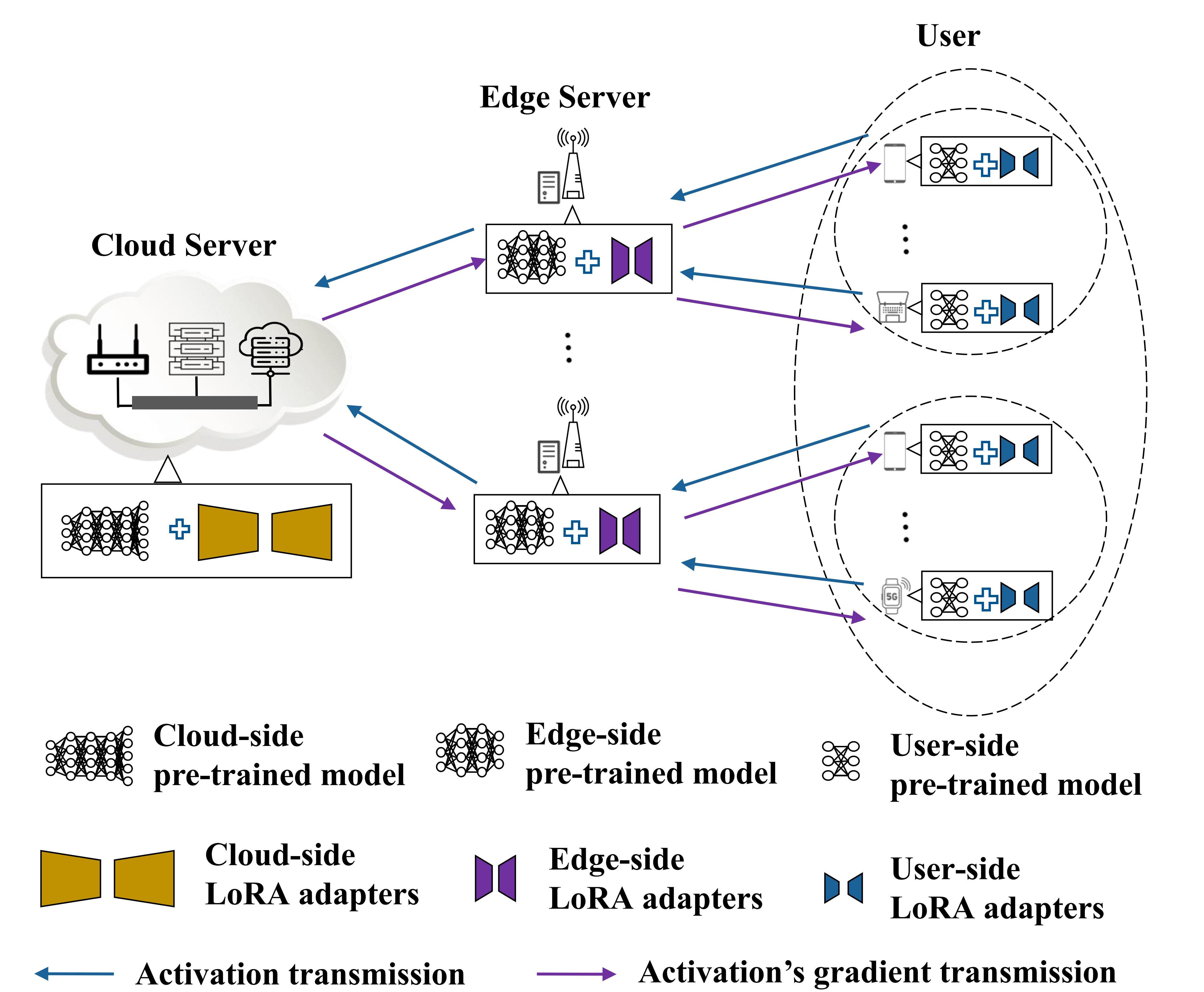}\label{SplitLLM}}\hfil
\subfloat[Lora-Based System]{\includegraphics[width=3.3in]{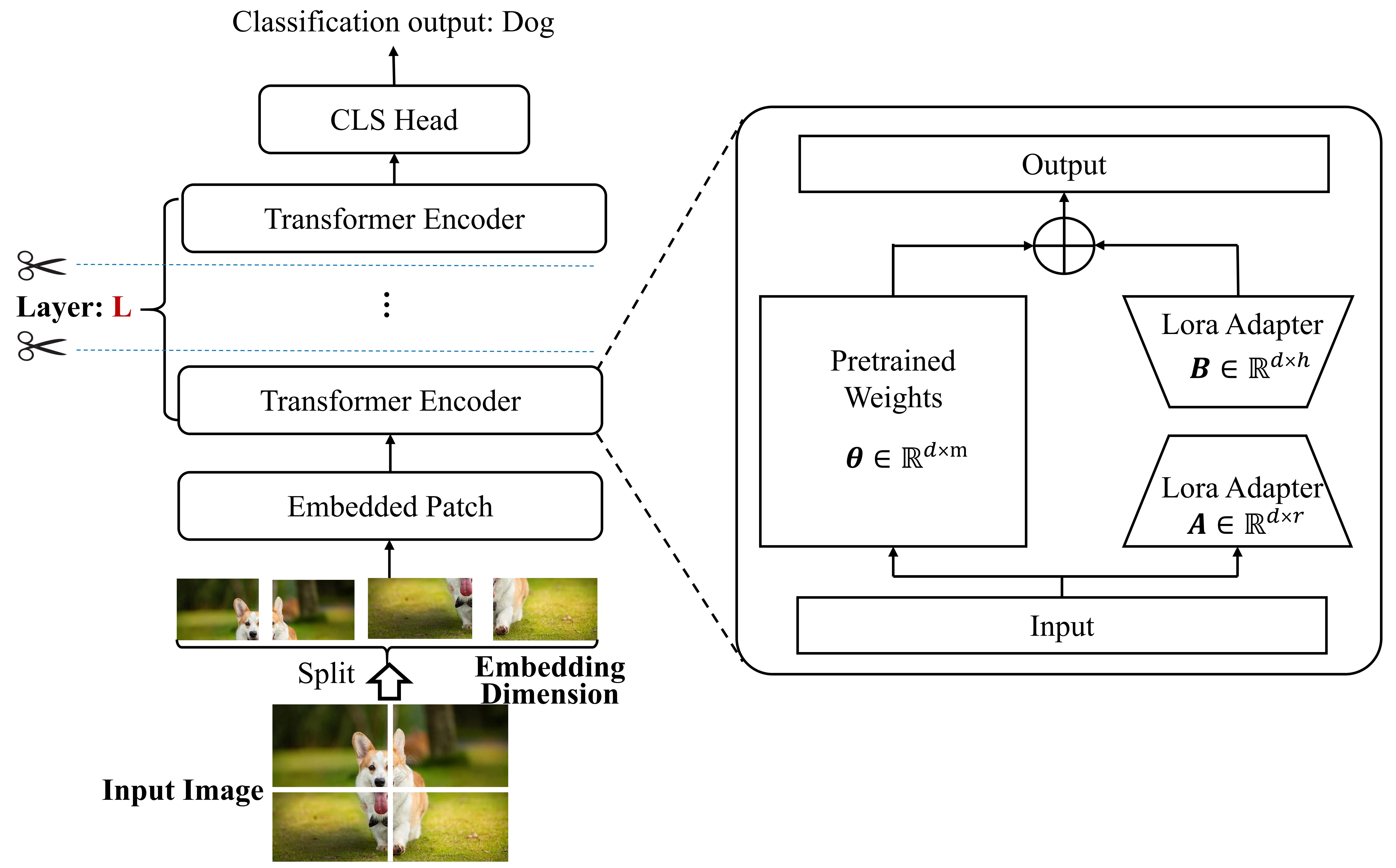}\label{CueLayer}}
\caption{(a) In the SplitLLM scheme, edge servers and corresponding users are trained parallelly$\mathrm{;}$ (b) in the LoRA-based learning system, only the LoRA adapter is updated, and the pre-trained model is divided into three components.}
\label{fig_cnn_performance}
\end{figure*}

In this paper, we propose a novel cloud-edge-user SL scheme for LLM fine-tuning, named SplitLLM. The proposed scheme reduces memory usage and computational workload by splitting the pre-trained model across cloud, edge, and users, and only updating the LoRA adapter.  Specifically, in each training round, users perform forward propagation based on local data and transmit the activation to the edge server, which continues to execute the forward propagation sequentially. The cloud server receives activation from the edge server and then executes the forward propagation. The cloud server then updates the cloud-side Lora adapter and returns the gradients to the corresponding edge servers and users to complete backward propagation. At the end of each training round, all users and edge servers upload their Lora adapter to the cloud server for aggregation, which is used for the next round of training. The proposed SplitLLM scheme offers three key advantages: (1) it enables training on large-scale user scenarios while reducing communication pressure on cloud servers, due to the three-layer scheme of user, edge, and cloud; (2) it alleviates memory pressure on both edge servers and cloud servers, as each component only handles a single pre-trained model; and (3) the split scheme enhances training efficiency, as the more critical model weights are frequently updated on the cloud server, similar to~\cite{freezing}. Experimental results demonstrate that our proposed scheme achieves comparable training accuracy, and reduces reduces peak memory usage up to $74\%$ compared to the state-of-the-art benchmarks.

The remainder of this paper is organized as follows. Section~II introduces the system model. Section~III describes the SplitLLM scheme.
Section~IV presents the simulation results. Section~V discusses the conclusion and future work.

\section{System Model}

\subsection{Considered Scenario}
In this section, we design an SL scheme across cloud, edge, and users for LLM, as shown in Fig. \ref{SplitLLM}. Specifically, we consider a wireless network scenario consisting of a cloud server, $M$ edge servers, and $N$ users, where $M << N$. We define the set of edge servers as $\mathcal{M}=\{1,\dots,M\}$ and the set of users as $\mathcal{N}=\{1,\dots,N\}$. In this scenario, the cloud server, with its superior computational and storage capabilities, is primarily responsible for the AI model in a fine-tuning training task and also manages model aggregation at each training round.
The edge server possesses intermediate computational capabilities, with only one edge-side model involved during the training process.
Each user possesses local data and a smaller AI model. All users' data contribute to a fine-tuning task, such as image classification, natural language processing, etc. 
In this cloud-edge-user scenario, we consider a fine-tuning task based on the Lora adapter, where the pre-trained large model and the Lora adapter are divided into three parts and deployed separately across the cloud, edge, and users.

\subsection{Lora-Based Learning Model}
We focus on supervised AI model training tasks. Let the training dataset of each user be defined as \(\mathcal{D}_n = \{(x_j, y_j)\}_{j=1}^{|\mathcal{D}_n|}\), where \(|\mathcal{D}_n|\) denotes the total number of samples for user \(n \in \mathcal{N}\). Here, \(x_j\) is the \(j\)-th input sample, and \(y_j\) is its corresponding label.
The vector \(\mathbf{\theta}\) represents the AI model parameter. 
The loss function for each sample, denoted as \(f_j(\mathbf{\theta})\), measures the prediction error for the \(j\)-th sample. The objective of the training process is to minimize the empirical loss \(F(\mathbf{\theta})\), which is defined as
\begin{equation}
   F(\mathbf{\theta}) = \frac{1}{|\mathcal{D}|} \sum_{j=1}^{|\mathcal{D}|} f(x_j, y_j; \mathbf{\theta}) = \frac{1}{|\mathcal{D}|} \sum_{j=1}^{|\mathcal{D}|} f_j(\mathbf{\theta}). 
\end{equation}
This process seeks to minimize the average prediction error across all samples and identify the optimal parameter $\mathbf{\theta}^*$.

% 为了能够支持更加高效的本地模型传输和聚合，我们在本地预训练的模型中插入一个小的低秩同构适配器。
To enhance the efficiency of local model transmission and aggregation, we integrate a small, low-rank, homogeneous adapter into the locally pre-trained model.
As depicted in Fig. \ref{CueLayer}, the LoRA adapter is designed to retain the same input dimension \(d\) and output dimension \(h\) as pre-trained large models, and is applicable to linear, embedding, and convolutional layers. Particularly for linear layers, the adapter reduces the parameter count by decomposing the conventional parameter matrix \( \mathbb{R}^{d \times h} \) into two smaller matrices \( \mathbf{A}\in\mathbb{R}^{d \times r} \) and \( \mathbf{B}\in\mathbb{R}^{r \times h} \), where the rank \( r \) is significantly lower than both \( d \) and \( h \). During the initialization phase, matrix \( \mathbf{A} \) is initialized with a Gaussian distribution with mean zero and variance \( \sigma^2 \), while matrix \( \mathbf{B} \) is set to zero~\cite{sun2024improving}. 
% 于是，我们设 \(W \in \mathbb{R}^{d \times m}\) 表示预训练模型的一个可训练权重矩阵，其相应的模型更新表示为 \(W + \Delta W = W + AB\)，其中 \(A \in \mathbb{R}^{d \times r}\)，\(B \in \mathbb{R}^{r \times m}\) 是分解矩阵，且 \(r \ll \min(d, m)\)。通过这种方式，可训练参数的数量可以比全参数微调减少超过99\%，从而提高了计算和通信效率。因此，我们在SplitLoRA框架中部署LoRA微调技术。
 In fine-tuning training, we define \(\mathbf{\theta} \in \mathbb{R}^{d \times m}\) as a trainable weight matrix of the pre-trained model, where the corresponding model update is expressed as \(\mathbf{\theta} + \Delta \mathbf{\theta} = \mathbf{\theta} + \mathbf{AB}\). 
 % Here, \(A \in \mathbb{R}^{d \times r}\) and \(B \in \mathbb{R}^{r \times m}\) serve as decomposition matrices, with \(r\) being significantly smaller than \(\min(d, m)\). 
 Lora adapter configuration allows the number of trainable parameters to be reduced significantly compared to full-parameter fine-tuning, thereby enhancing computational and communication efficiency. Accordingly, we deploy the LoRA adapter within the SplitLLM scheme.
Thus, the objective of SplitLLM can be rewritten as 
\begin{equation}\label{loss function}
    \begin{aligned}
        F(w) = \frac{1}{|D|} \sum_{j=1}^{|D|} f(x_j, y_j; \mathbf{\theta} + \mathbf{AB}) = \frac{1}{|D|} \sum_{j=1}^{|D|} f_j(\mathbf{\theta} + \mathbf{AB}).
    \end{aligned}
\end{equation}
In fine-tuning training, the aim is to minimize the average prediction error across all samples and identify the optimal parameter $\mathbf{A}^*\mathbf{B}^*$.

\begin{algorithm}[t]
\SetAlgoLined
%\KwIn{$\epsilon_u,~\epsilon_e,~$\epsilon_c$,~M, \mathrm{and} ~N$;}
%\KwOut{\mathbf{A^*B^*};}
Initialize the cut layer and Lora adapter parameter;\\
Cloud server broadcasts the edge-side and user-side pre-trained models;\\
\For{training round $t = 1,\ldots,T$}{
    Cloud server broadcasts the latest edge-side Lora adapter and user-side Lora adapter, respectively;\\
    \For{each edge server in parallel}{
        \For{user $n_1,\ldots,N_m$ in edge server $m$}{
            \For{local epoch $k=1,\dots,K$}{
                Randomly selects a mini-batch sample from the local data;\\
                Execute user-side pre-trained model and Lora adapter, then obtain an immediate result;\\
                Transmit immediate result to the edge server;\\
                \(\triangleright\) Edge server $m$ \textbf{execute}\\
                {
                    Receives the immediate result from the user;\\
                    Executes the edge-side pre-trained model and Lora adapter, then obtains immediate result;\\
                    Transmit immediate result to the cloud server;\\
                    \(\triangleright\) Cloud server \textbf{execute}\\}{
                        Receives the immediate result from edge server;\\
                        Executes the cloud-side pre-trained model and Lora adapter;\\
                        Updates the server-side Lora adapter;\\
                        Transmits the edge-side Lora adapter's gradient;\\
                    }
                    Updates the edge-side Lora adapter;\\
                    Transmits the user-side Lora adapter's gradient;\\
                }
            }
            Update the user-side Lora adapter to the edge server;\\
        }
        Sends the latest user-side Lora adapter to the edge server;\\
    }
    Sends the latest user-side and edge-side Lora adapters to the Cloud server;\\

Cloud aggregates edge-side Lora adapters and user-side Lora adapters into a new one;\\
\caption{Cloud-Edge-User Hierarchical SplitLLM.}
\label{Split algorithm}
\end{algorithm}

\section{Cloud-Edge-User Hierarchical SplitLLM}
The proposed scheme primarily comprises three steps: model distribution, model training, and model aggregation. The objective of model distribution is to distribute the split pre-trained model and Lora adapter to edge servers and users. Model training involves multiple epochs of execution and updates, aimed at completing both forward and backward propagation. Finally, model aggregation is performed on the cloud server, where the well-trained Lora adapters from different users and edge servers are aggregated into new adapters. The details of the SplitLLM are presented in Alg.~\ref{Split algorithm}.

%\vspace{-0.1cm}
\subsubsection{Step 1 - Model Distribution}

In the initial phase of training, the pre-trained model, which consists of $L$ layers, is divided into three parts: $\mathbf{\theta}_u$ for the user-side model, $\mathbf{\theta}_e$ for the edge server model, and $\mathbf{\theta}_s$ for the cloud server model. 
The segmented pre-trained model is distributed layer-by-layer from the cloud to the edge and then to the user, and this distribution occurs only once throughout the training process because the pre-trained model does not require updates and aggregation in each training round.
We assume a total of $L$ Lora adapters, denoted as $\mathcal{L}=\{1,\dots,L\}$.
Specifically, the user model consists of the first layer, denoted as $\mathcal{L}_u=\{1\}$. The edge server consists of layers 2 through $L_e$, represented by $\mathcal{L}_e=\{2,\dots,L_e\}$, while the cloud server handles the remaining layers from $L_e+1$ to $L$, indicated as $\mathcal{L}_c=\{L_e+1,\dots,L\}$.
We assume the user-side LoRa adapter, $\mathbf{\eta}_u$, consists of $\{\mathbf{A}^1,\mathbf{B}^1\}$; the edge server's adapter, $\mathbf{\eta}_e$, includes $\{\mathbf{A}^2, \mathbf{B}^2, \ldots, \mathbf{A}^{L_e}, \mathbf{B}^{L_e}\}$; and the cloud server's adapter, $\mathbf{\eta}_s$, comprises $\{\mathbf{A}^{L_e+1}, \mathbf{B}^{L_e+1}, \ldots, \mathbf{A}^L, \mathbf{B}^L\}$.Similarly, the adapters $\mathbf{\eta}_u$ and $\mathbf{\eta}_e$ are also distributed from the cloud to the edge and to the user.

% %\vspace{-0.1cm}
\subsubsection{Step 2 - Model Training} This process involves performing forward propagation to obtain prediction results and using backward propagation to update the adapters on the cloud server, edge servers, and users. Each training round executes $K$ training epochs, and the entire model training process is divided into the model execution and the model update phase.

\begin{itemize}
\item \textbf{Model execution:~}

Assume the training spans $T$ rounds. In round $t$, the $n^{th}$ user under the $m^{th}$ edge server randomly selects a mini-batch from their local data, denoted by $\mathcal{B}_{m,n}(t,k) \subseteq \mathcal{D}_{m,n}$, to perform forward propagation. Here, $B = |\mathcal{B}_{m,n}(t,k)|$ represents the size of each mini-batch, and $\mathcal{D}_{m,n}$ is the local dataset of user $n$, where $k$ denotes the number of epochs the user undergoes in local training. The forward propagation involves passing $\mathcal{B}_{m,n}(t,k)$ through $\mathbf{\theta}_u(t,k)$ and $\mathbf{\eta}_u$ to produce the output $R_{m,n}^u(t,k)$. This process can be mathematically represented as
\begin{equation}
\mathbf{R}_{m,n}^u(t,k) = f(\mathbf{\theta}_u(t,k), \mathbf{\eta}_u(t,k), \mathcal{B}_{m,n}(t,k))
\end{equation}

All users transmit their locally computed outputs to edge server $m$. The edge server then inputs $\mathbf{R}_{m,n}^u(t,k)$ into its model, using parameters $\mathbf{\theta}_e(t,k)$ and $\mathbf{\eta}_e(t,k)$, to continue the forward propagation training. This process can be mathematically represented as
\begin{equation}
\mathbf{R}_{m,n}^e(t,k) = f(\mathbf{\theta}_e(t,k), \mathbf{\eta}_e(t,k), \mathcal{B}_{m,n}(t,k))
\end{equation}

The edge server forwards its output results to the cloud server, which then conducts forward propagation using the cloud server model parameters $\mathbf{\theta}_c(t,k)$ and $\mathbf{\eta}_c(t,k)$. This process can be mathematically represented as
\begin{equation}
\mathbf{y}_{m,n}^c(t,k) = f(\mathbf{\theta}_s(t,k), \mathbf{\eta}_s(t,k), \mathcal{R}_{m,n}^e(t,k))
\end{equation}
\end{itemize}

\begin{itemize}
\item \textbf{Model updating:~} 
% In the process of updating the model, updates are required for both the server-side and the client-side models. Specifically, the AP can compute the loss function and gradient of each mini-batch data based on the prediction results obtained through forward propagation. The update of the server-side model parameters can be accomplished through methods such as stochastic gradient descent. The completion of the above calculation process by the AP not only updates the model, but also performs the backward propagation of the model on the server-side. Then, the gradient of the smashed data will be transmitted to the corresponding client in the group. Upon receiving the gradient of the smashed data, each client-side model is updated locally.
% 后向传播的目的是为了最小化全局损失函数\eqref{loss function}。首先是云服务基于预测结果和标签计算云服务器、边缘服务器以及用户中Lora的梯度$g_{m,n}^{A,l}(t,k)，~i\in{1,\dots,L}$和$g_{m,n}^{B,l}(t,k)，~i\in{1,\dots,L}$。随后云服务器会首先更新自身Lora模型的参数
% \begin{equation}
%     \begin{aligned}
%     \mathbf{A}_{m,n}^{l}{t,k} \leftarrow \mathbf{A}_{m,n}^{l}{t-1,k}  - \epsilon_s g_{m,n}^{A,l}(t,k)，~i\in{L_e+1,\dots,L}
%     \end{aligned}
% \end{equation}
% \begin{equation}
%     \begin{aligned}
%     \mathbf{B}_{m,n}^{l}{t,k} \leftarrow \mathbf{B}_{m,n}^{l}{t-1,k}  - \epsilon_s g_{m,n}^{B,l}(t,k)，~i\in{L_e+1,\dots,L}
%     \end{aligned}
% \end{equation}
During this process, the model updates require updating the Adapter parameters across the cloud server, edge servers, and users, which completes the backward propagation in AI model training.
The purpose of backward propagation is to minimize the global loss function as defined in \eqref{loss function}. The cloud server calculates gradients for both the cloud-side, edge-side, and user-side LoRa adapter $g_{m,n}^{A,l}(t,k)$ and $g_{m,n}^{B,l}(t,k)$, where $l \in \mathcal{L}$, based on the prediction outcomes $\mathbf{y}_{m,n}^c(t,k)$ and labels. Subsequently, the cloud server updates clod-side LoRa adapter parameters as follows:
\begin{equation}
    \mathbf{A}_{m,n}^l(t,k) \leftarrow \mathbf{A}_{m,n}^l(t-1,k) - \epsilon_s g_{m,n}^{A,l}(t,k), \forall l \in \mathcal{L}_c,
\end{equation}
and
\begin{equation}
    \mathbf{B}_{m,n}^l(t,k) \leftarrow \mathbf{B}_{m,n}^l(t-1,k) - \epsilon_s g_{m,n}^{B,l}(t,k), \forall l \in \mathcal{L}_c
\end{equation}
% 其中$\epsilon_s$是云服务上lora adapter的学习率。为了简化，我们假设每一个encoder block是链式的，lora adapter是从最后一个更新到cut layer层，也即是第$L_e+1$层。当梯度更新到第$L_e+1$层时，$1$到$L$层Lara的梯度会传回到对应的边缘服务器。边缘服务器进而更新边缘服务端的Lora adapter
% \begin{equation}
%     \mathbf{A}_{m,n}^l(t,k) \leftarrow \mathbf{A}_{m,n}^l(t-1,k) - \epsilon_e g_{m,n}^{A,l}(t,k), \forall l \in \{2, \dots, L_e\}
% \end{equation}
% \begin{equation}
%     \mathbf{B}_{m,n}^l(t,k) \leftarrow \mathbf{B}_{m,n}^l(t-1,k) - \epsilon_e g_{m,n}^{B,l}(t,k), \forall l \in \{2, \dots, L_e\}
% \end{equation}
% 其中$\epsilon_e$是边缘服务上lora adapter的学习率。随后云服务器发送第$1$层的Lora adapter梯度给本地用户，本地用户进而更新本地的lora adapter参数：
% \begin{equation}
%     \mathbf{A}_{m,n}^l(t,k) \leftarrow \mathbf{A}_{m,n}^l(t-1,k) - \epsilon_u g_{m,n}^{A,l}(t,k), \forall l \in \{1\}
% \end{equation}
% \begin{equation}
%     \mathbf{B}_{m,n}^l(t,k) \leftarrow \mathbf{B}_{m,n}^l(t-1,k) - \epsilon_u g_{m,n}^{B,l}(t,k), \forall l \in \{1\}
% \end{equation}

where $\epsilon_s$ denotes the learning rate for the LoRa adapter on the cloud server. For simplicity, assume each encoder block is linked sequentially, with the LoRa adapter updating from the last to the cut $(L_e+1)^{th}$ layer. When the gradient is updated to the cut layer, gradients for layers $1$ through $L$ of LoRa adapters are transmitted back to the corresponding edge servers. The edge servers then update the edge-side LoRa adapters as follows:
\begin{equation}
    \mathbf{A}_{m,n}^l(t,k) \leftarrow \mathbf{A}_{m,n}^l(t-1,k) - \epsilon_e g_{m,n}^{A,l}(t,k), \forall l \in \mathcal{L}_e,
\end{equation}
and
\begin{equation}
    \mathbf{B}_{m,n}^l(t,k) \leftarrow \mathbf{B}_{m,n}^l(t-1,k) - \epsilon_e g_{m,n}^{B,l}(t,k), \forall l \in \mathcal{L}_e
\end{equation}
where $\epsilon_e$ is the learning rate for the LoRa adapter on the edge server. Subsequently, the edge server sends the LoRa adapter gradients for the first layer to the local user, who then updates user-side LoRa adapter parameters, which can be represented as
\begin{equation}
    \mathbf{A}_{m,n}^l(t,k) \leftarrow \mathbf{A}_{m,n}^l(t-1,k) - \epsilon_u g_{m,n}^{A,l}(t,k), \forall l \in \mathcal{L}_e,
\end{equation}
and
\begin{equation}
    \mathbf{B}_{m,n}^l(t,k) \leftarrow \mathbf{B}_{m,n}^l(t-1,k) - \epsilon_u g_{m,n}^{B,l}(t,k), \forall l \in \mathcal{L}_u
\end{equation}

\end{itemize}

\subsubsection{Step 3 - Model Aggregation}
%\vspace{-0.1cm}
% After all groups have completed the model training process and generated new whole models on the AP, AP will aggregate these new whole models into a single model to complete an epoch of training. The process of model aggregation can be accomplished through algorithms such as FedAVG.
% The aggregated model will be involved in new epochs of model training and model aggregation until the target number of training epochs is reached.
% 用户在完成本地lora adapter的更新之后，则是完成了一个epoch的训练。在所有用户完成$K$轮的训练之后，则进入了模型聚合过程。所有用户会把自身更新后的lora adapter上传到对应的边缘服务器上，边缘服务进而转发上传到云服务器。同时边缘服务器会把边缘端的Lora adapter上传到云服务器做聚合。云服务器通过Fedaverage的方法将这些adapter进行聚合。用户们的聚合过程可以表示为
% \begin{equation}
%     \begin{aligned}
%     \mathbf{A}_{m,n}^l(t+1,1) = \sum_{n=1}^N \sum_{m=1}^M \frac{|D_{m,n}|}{|D|}     \mathbf{A}_{m,n}^l(t,K),  
%     \end{aligned}
% \end{equation}
% \begin{equation}
%     \begin{aligned}
%     \mathbf{B}_{m,n}^l(t+1,1) = \sum_{n=1}^N \sum_{m=1}^M \frac{|D_{m,n}|}{|D|}     \mathbf{B}_{m,n}^l(t,K),  
%     \end{aligned}
% \end{equation}
After completing the update of the local LoRa adapter, a user concludes an epoch of training. Following the completion of $K$ training rounds by all users, the model aggregation process begins. Each user uploads their updated LoRa adapter to the corresponding edge server, which then forwards the uploads to the cloud server. Additionally, the edge servers upload their respective edge-side LoRa adapters to the cloud server for aggregation. The cloud server aggregates these adapters using the FedAvg method. The aggregation process can be mathematically represented~as
\begin{equation}
    \mathbf{A}_{m,n}^l(t+1,1) = \sum_{n=1}^N \sum_{m=1}^M \frac{|D_{m,n}|}{|D|} \mathbf{A}_{m,n}^l(t,K),~ \forall l \in \mathcal{L},
\end{equation}
and 
\begin{equation}
    \mathbf{B}_{m,n}^l(t+1,1) = \sum_{n=1}^N \sum_{m=1}^M \frac{|D_{m,n}|}{|D|} \mathbf{B}_{m,n}^l(t,K),~ \forall l \in \mathcal{L}.
\end{equation}
The aggregated adapter will be involved in a new training round until the target accuracy is reached.

% \begin{figure}[!t]
% \centering
% \begin{adjustbox}{max width=\textwidth}
% \begin{tabular}{@{}cc@{}}
% \subfloat[Accuracy]{\includegraphics[width=0.5\linewidth]{./figure2.eps}%
% \label{case1}}\hspace{-4mm} &
% \subfloat[Delay]{\includegraphics[width=0.5\linewidth]{./figure3.eps}%
% \label{case2}}\\[-2mm]
% \end{tabular}
% \end{adjustbox}
% \caption{Performance comparison with respect to training accuracy and training latency on the GTSRB dataset.}
% %\vspace{-0.4cm}
% \label{fig_sim}
% \end{figure}
\begin{figure*}[htbp]
\centering
\subfloat[Performance of CIFAR100 on ViT Model under IID Conditions]{\includegraphics[width=2.9in]{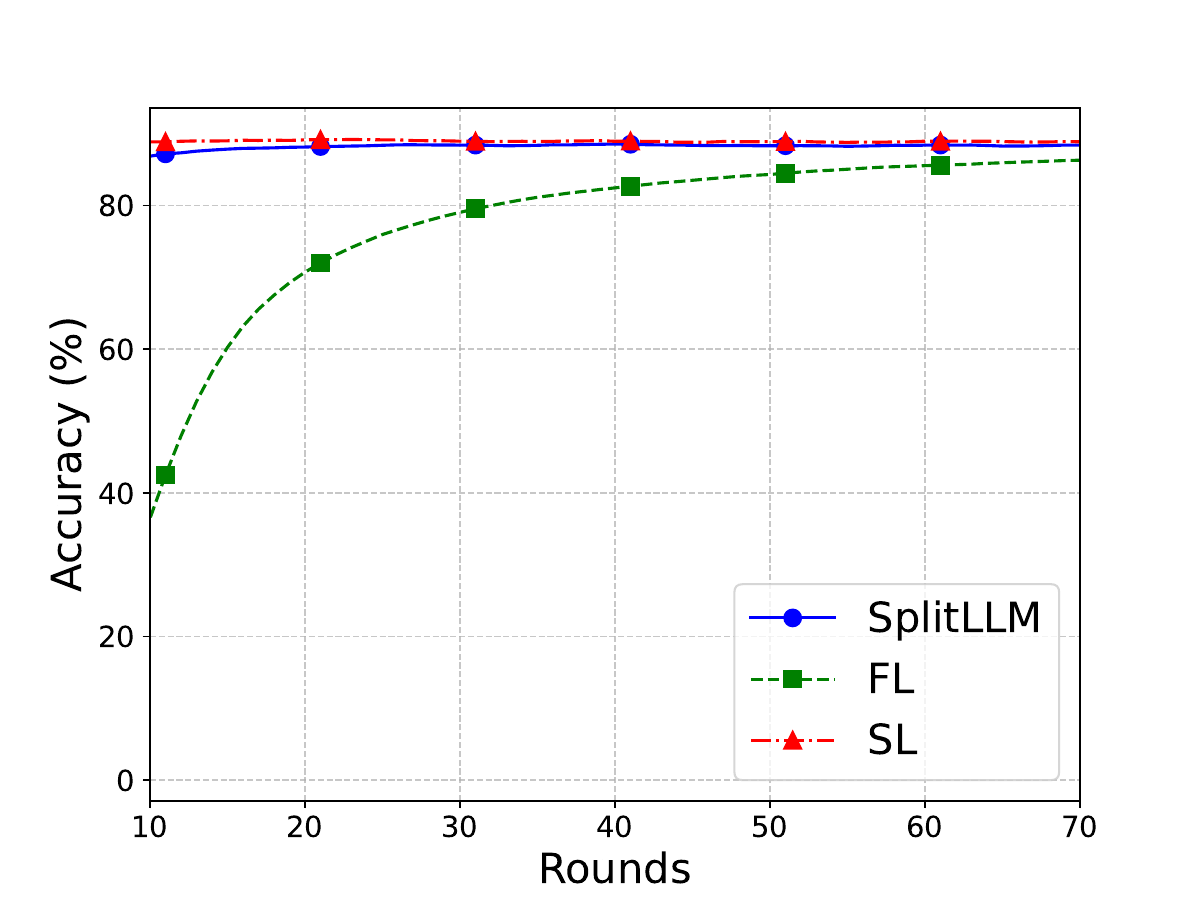}\label{GHSFLSCHEME}}\hfil
\subfloat[Performance of CIFAR100 on ViT Model under non-IID Conditions]{\includegraphics[width=2.9in]{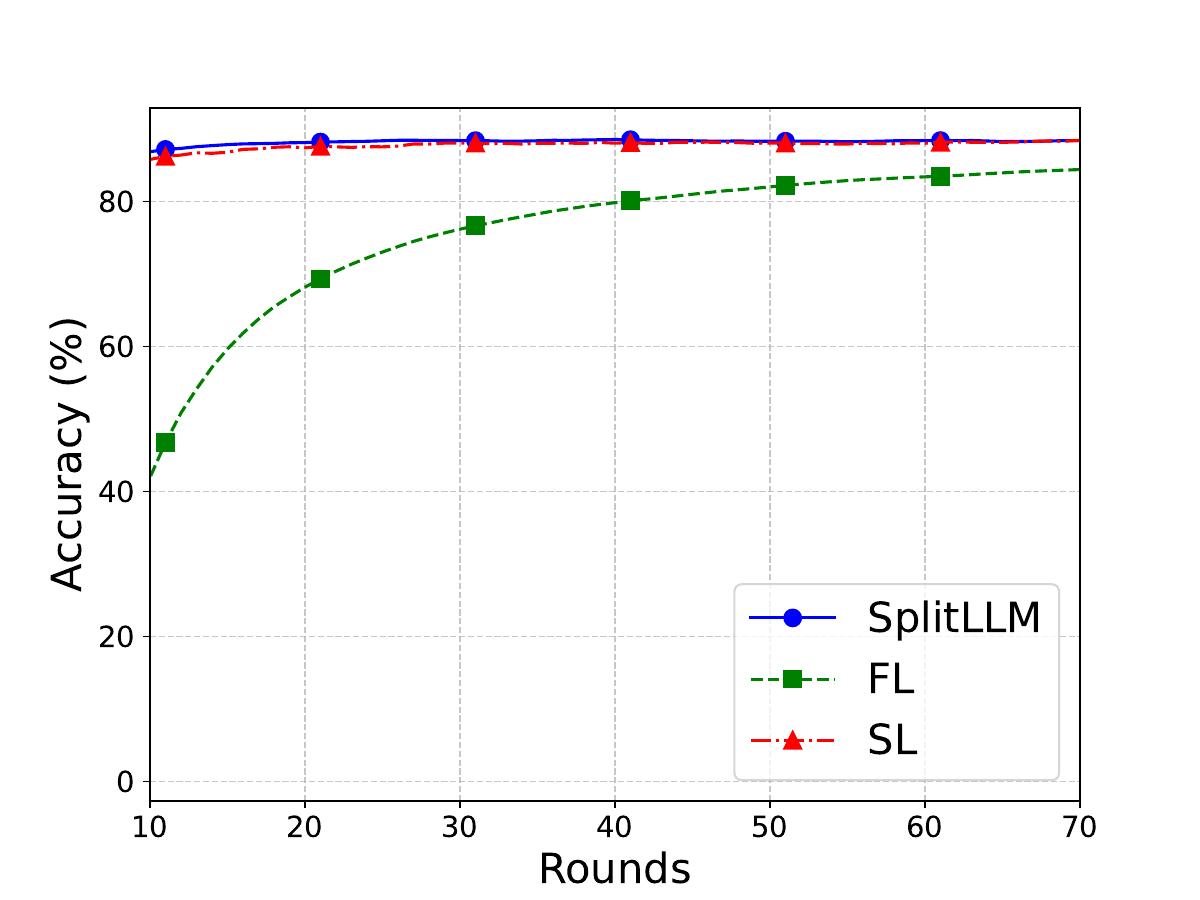}\label{SLSCHEME}}\\
\subfloat[Performance of MRPC on BERT Model under IID Conditions]{\includegraphics[width=2.9in]{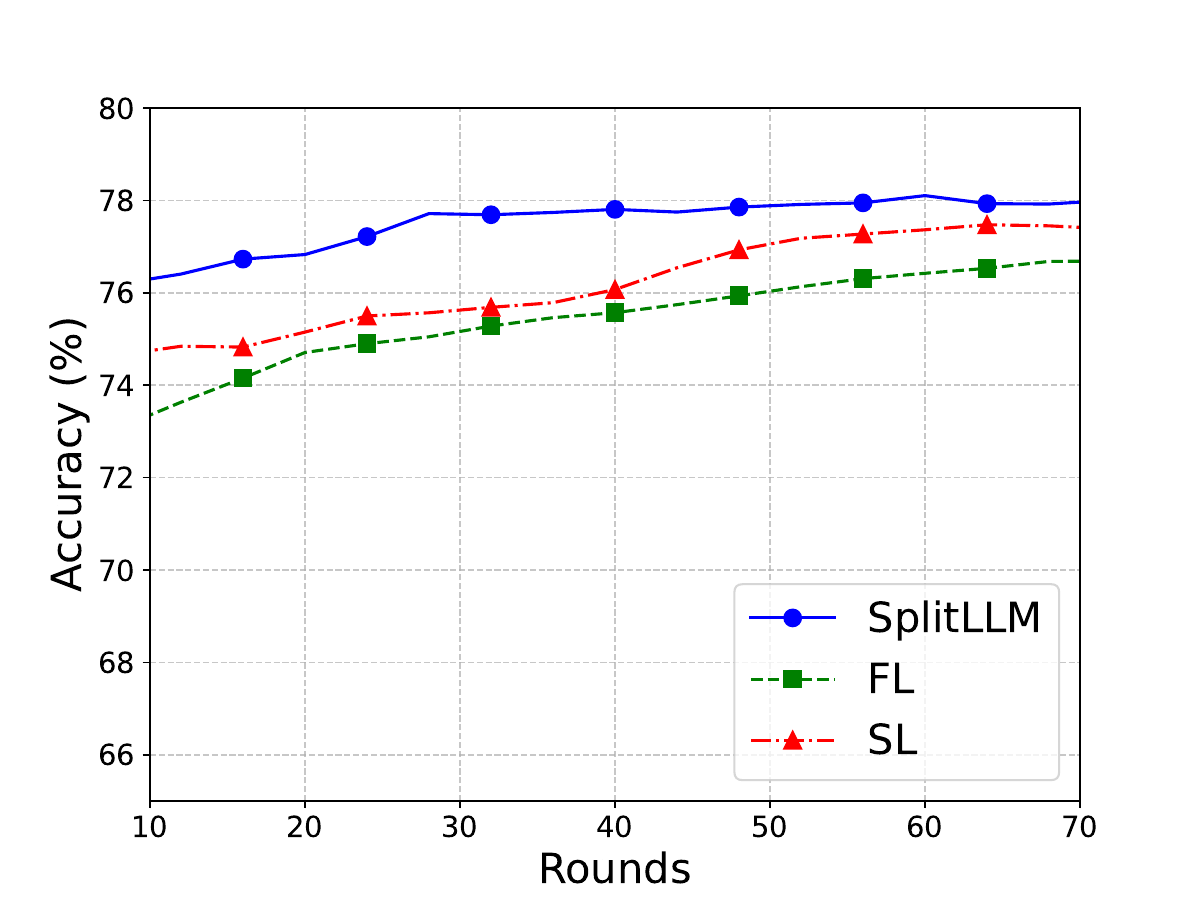}\label{SLSCHEME}}\hfil
\subfloat[Performance of MRPC on BERT Model under non-IID Conditions]{\includegraphics[width=2.9in]{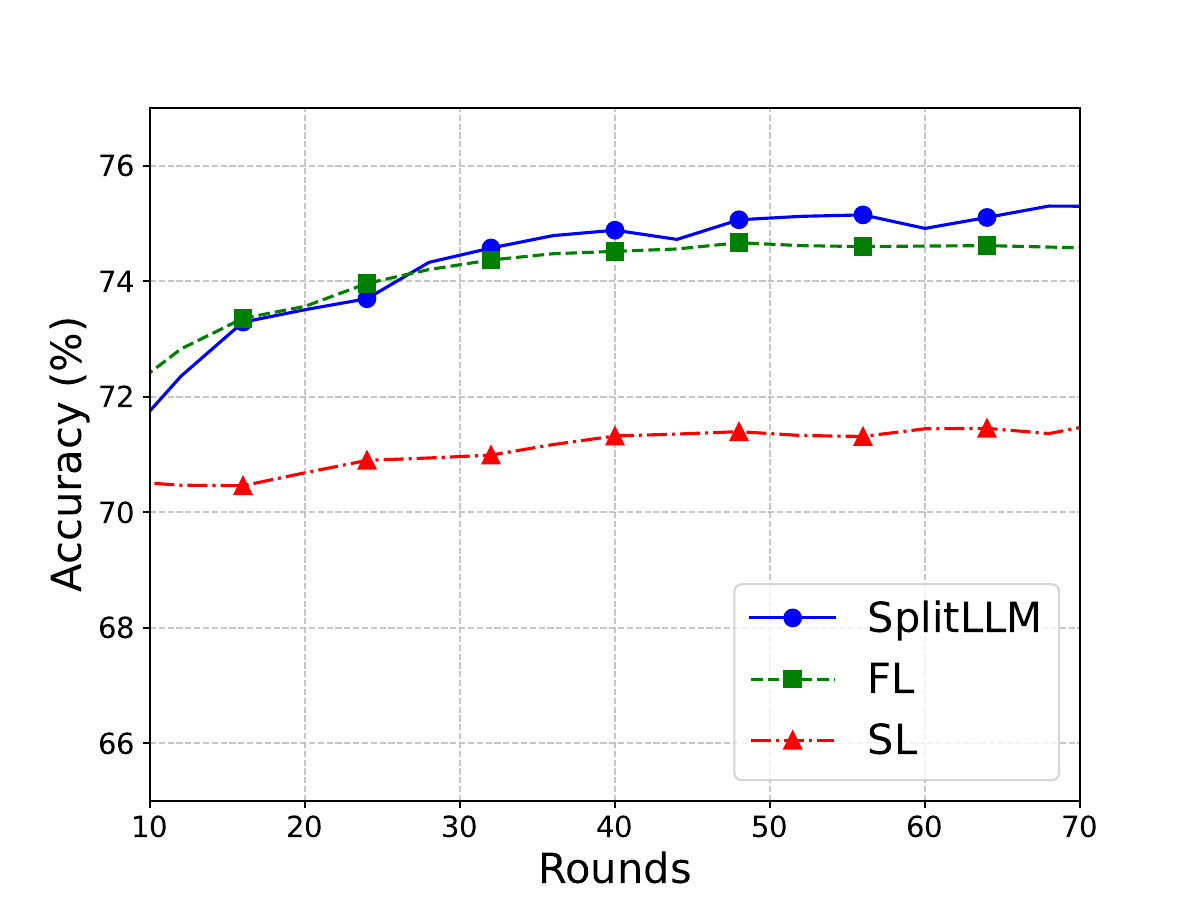}\label{SLSCHEME}}
\caption{Performance on various neural architectures and datasets.}
\label{fig:four_figures}
\end{figure*}

\section{Simulation Results}
\subsection{Experiment Settings}
For the experimental evaluations, we focus on the image classification task using the CIFAR100 dataset \cite{krizhevsky2009learning} and the semantic similarity prediction task using the MRPC dataset~\cite{wang2018glue}. We adopt ViT-Base and BERT-Base as the pre-trained vision and language models, respectively. We then apply LoRA to fine-tune all the linear and convolutional layers in these two pre-trained models. The details of the hyperparameter settings for model training are presented in Table \ref{tab:hyperparam}. In the FL and SL setup, we deploy a total of 20 users and a cloud server. In our proposed SplitLLM, we also deploy 5 edge servers as intermediate nodes with computing capacity. For the independent and identically distributed (IID) setting, we randomly partition the dataset into multiple shards and uniformly assign them to each user. For the non-IID setting, we use the Dirichlet distribution with a concentration parameter of 0.5 for the dataset partition \cite{wang2020federated}.

\begin{table}[t]
\centering
\caption{The hyperparameters for model training}
\label{tab:hyperparam}
\setlength{\tabcolsep}{5mm}{
\begin{tabular}{rcc}
\toprule
\multicolumn{1}{l}{} & CIFAR100 & MRPC      \\ \hline
\#training samples   & 50,000    & 3668      \\
\#test samples       & 10,000    & 1725      \\
model                & ViT-base & bert-base \\
\#model parameters   & 86M      & 110M      \\
optimizer            & SGD      & AdamW     \\
learning rate        & 1e-4     & 2e-5      \\
momentum             & 0.9      & -         \\
lr\_decay            & 0.998    & 0.998     \\
batchsize            & 32       & 16        \\
local epoch          & 1        & 1         \\
lora rank            & 8        & 8         \\ 
\bottomrule
\end{tabular}
}
\end{table}

\subsection{Performance Evalutions}
We next compare the proposed SplitLLM with the following two distributed learning frameworks:
(1) FL \cite{mcmahan2017communication}: We employ a vanilla FL framework, where the entire model is trained locally on edge users and then transmitted to the server for model synchronization.
(2) SL \cite{wu2023split}: We employ a SL framework, where the user-side model is sequentially trained across users in collaboration with a shared server-side model. The user and cloud server transmit the intermediate layer outputs and gradients, respectively, to collaboratively train the entire model.

Fig.~\ref{fig:four_figures} illustrates the learning performance of different methods across various neural architectures and datasets. The proposed SplitLLM outperforms the baseline methods in improving training efficiency in most cases. Additionally, SplitLLM adapts well to different types of datasets and data distribution settings. In small sample sets and extreme data distribution setting, such as the MRPC task in a non-iid setting, the model training does not exhibit any significant performance degradation like SL.

Table \ref{tab:performance} presents the comparison results of SplitLLM against the baseline methods in terms of best accuracy, user-side communication cost, and peak memory usage across user, edge, and cloud. The proposed SplitLLM reduces peak memory usage by $74\%$ compared to traditional FL while maintaining high learning performance. Furthermore, one of the key advantages of SplitLLM is its robust learning capacity, which demonstrates greater resilience than SL in small sample sets and non-iid settings.

\begin{table}[t]
\scriptsize
\centering
\caption{Performance comparison between SplitLLM\\ and other frameworks}
\label{tab:performance}
\setlength{\tabcolsep}{1mm}{
\begin{tabular}{ccccc}
\toprule
Dataset & Framework & \begin{tabular}[c]{@{}c@{}}Best Acc.(\%)\\ iid/non-iid\end{tabular} & \begin{tabular}[c]{@{}c@{}}User-Side\\ Comm. Cost(GB)\end{tabular} & \begin{tabular}[c]{@{}c@{}}User/Edge/Cloud\\ Memory Usage(GB)\end{tabular} 
\\ 
\hline
 \multirow{3}{*}{MRPC} & SplitLLM   & 78.23/75.63  & 0.1289  & 1.39/1.71/2.25 
\\
 & FL & 77.49/77.34  & 0.0099  & 5.35/----/----  
\\
& SL  & 77.82/71.84  & 0.1289  & 1.39/----/3.96 
\\
\hline
 \multirow{3}{*}{CIFAR100} & SplitLLM   & 89.64/88.77  & 2.81  & 1.56/1.98/3.76 
\\
& FL   & 87.66/86.02  & 0.0089  & 7.21/----/---- 
\\
& SL  & 89.81/87.87  & 2.81  & 1.56/----/5.75 
\\ 
\bottomrule
\end{tabular}
}
\end{table}
\section{Conclusion}
In this paper, we have proposed a cloud-edge-user-based SL scheme for fine-tuning LLM. The proposed scheme divides the pre-trained model and LoRA adapters into cloud-side, edge-side, and user-side parts, with only LoRA adapters updated. During the training, forward propagation proceeds from the user to the edge and then to the cloud, while backward propagation follows the reverse order. Experimental results show that compared to cloud-based FL and SL, the proposed scheme significantly reduces both memory load and communication overhead. The proposed fine-tuning scheme is suitable for wireless networks with a large number of users and limited memory resources on the server. In future work, we will investigate the cut layer selection problem to balance the trade-off between computational workload and communication overhead under devices' memory constraints.
%\vspace{-0.1cm}
% In this paper, we have validated that the effectiveness of our proposed scheme can greatly reduce training latency while preserving high model accuracy performance through experiments. For future work, we will study the impact of a cut layer selection and client grouping on the system performance. Meanwhile, the rational allocation of communication bandwidth resources and computing resources is crucial for further enhancing the performance, and hence we will design an efficient and effective resource allocation scheme for the proposed scheme. 

%\vspace{-0.1cm}
% \section*{ACKNOWLEDGEMENT}
% This work was supported in part by the Peng Cheng Laboratory Major Key Project under Grant PCL2021A09-B2 and by the Natural Science Foundation of China under Grant 6220012314.
\section*{Acknowledgment}
This work was supported in part by the Peng Cheng Laboratory Major Key Project under Grants PCL2023AS1-5 and PCL2021A09-B2, in part by the Natural Science Foundation of China under Grant 62201311, and in part by the Young Elite Scientists Sponsorship Program by CAST under Grant 2023QNRC001.

\bibliographystyle{IEEEtran}
%\vspace{-0.1cm}

\bibliography{Globecom.bib}

\end{document}